\documentclass[aps,pra,reprint,amsmath,amssymb,%
   twocolumn,showpacs,showkeys,floatfix
]{revtex4-1}

\usepackage{bm}
\usepackage{graphicx}
\usepackage{hyperref}
\usepackage[english]{babel}
\usepackage[usenames,dvipsnames,svgnames,table]{xcolor}
\usepackage[color,final]{showkeys}
\definecolor{refkey}{wave}{623} 
\definecolor{labelkey}{wave}{423} 

\newcommand{\mycomment}[1]{}

\newcommand{\ket}[1]{\ensuremath{|{#1}\rangle}}
\newcommand{\bra}[1]{\ensuremath{\langle{#1}|}}
\newcommand{\ketbra}[1]{\ensuremath{|{#1}\rangle\langle{#1}|}}

\newcommand{\JzeroSpin}{g}
\newcommand{\JtwoSpin}{e}
\newcommand{\ground}{\ensuremath{|\JzeroSpin\rangle}}
\newcommand{\excit}{\ensuremath{|\JtwoSpin\rangle}}
\newcommand{\ztoway}{\ensuremath{z_{\uparrow,\downarrow}}}
\newcommand{\ztow}{\ensuremath{z_{\uparrow}}}
\newcommand{\zaway}{\ensuremath{z_{\downarrow}}}
\newcommand{\Vtow}{\ensuremath{V_{\uparrow}}}
\newcommand{\Vaway}{\ensuremath{V_{\downarrow}}}

\newcommand{\Brot}{\ensuremath{B_{\mathrm{rot}}}}
\newcommand{\Trot}{\ensuremath{T_{\mathrm{rot}}}}

\newcommand\mycalS{\ensuremath{{\cal S}}}
\newcommand\mycalD{\ensuremath{{\cal D}}}
\newcommand{\dip}{\ensuremath{\mu}}
\newcommand{\dipvec}{\ensuremath{\bm{\mu}}}
\newcommand{\dipz}{\ensuremath{\mu_z}}
\newcommand{\dipmol}{\ensuremath{\mu_{\mathrm{mol}}}}
\newcommand{\om}{\omega}
\newcommand{\Om}{\Omega}
\newcommand{\half}{\ensuremath{\frac{1}{2}}}

\newcommand{\com}{\textrm{c.m.}}
\newcommand{\str}{\textrm{str}}

\newcommand{\Ca}[0]{\ensuremath{{^{40}\mathrm{Ca}^+}}}
\newcommand{\CaH}[0]{\ensuremath{\mathrm{CaH}}}
\newcommand{\CaHgs}[0]{\CaH~\ensuremath{(X {}^2\Sigma)}}
\newcommand{\KRb}[0]{\ensuremath{^{40}\mathrm{K}^{87}\mathrm{Rb}}}
\newcommand{\KRbgs}[0]{\ensuremath{^{40}\mathrm{K}^{87}\mathrm{Rb}~(X {}^1\Sigma^{+}\!)}}

\newcommand{\omcomstr}{\om_{\com,\str}}
\newcommand{\omcom}{\om_{\com}}

\newcommand{\omstr}{\om_{\str}}
\newcommand{\omstrz}{\om_{\str,0}}

\newcommand{\mred}{\ensuremath{m_{\mathrm{red}}}}
\newcommand{\mtot}{\ensuremath{m_{\mathrm{tot}}}}
\newcommand{\talpha}{\ensuremath{\tilde{\alpha}}}

\newcommand{\erot}{\ensuremath{E_{\mathrm{rot}}}}
\newcommand{\mumeter}[0]{\ensuremath{\mu\mathrm{m}}}
\newcommand{\rvec}{\ensuremath{\bm{r}}}
\newcommand{\rdip}{\ensuremath{r_{\mathrm{dip}}}}

\newcommand{\R}{\bm{R}}
\newcommand{\z}{\bm{z}}
\newcommand{\x}{\bm{x}}
\newcommand{\X}{\bm{X}}

\begin{document}

\title{Measuring molecular electric dipoles using trapped atomic ions and ultrafast laser pulses}

\author{Jordi Mur-Petit}
\email[Corresponding author: ]{jordi.mur@csic.es}
\affiliation{Instituto de Estructura de la Materia, IEM-CSIC, Serrano 123,
  E-28006 Madrid, Spain}
\affiliation{Kavli Institute for Theoretical Physics, University of California, Santa Barbara, CA 93106, USA}
\author{Juan Jos\'e Garc\'\i a-Ripoll}
\affiliation{Instituto de F\'\i sica Fundamental, IFF-CSIC, Serrano 113 bis,
  E-28006 Madrid, Spain}



\preprint{NSF-KITP-13-103}

\begin{abstract}
We study a hybrid quantum system composed of an ion and an electric dipole.
We show how a trapped ion can be used to measure the small electric field generated by a classical dipole.
We discuss the application of this scheme to measure the electric dipole moment of cold polar molecules, whose internal state can be controlled with ultrafast laser pulses, by trapping them in the vicinity of a trapped ion.
\end{abstract}

\maketitle


\section{Introduction}\label{sec:intro}

Outstanding progress during the last 20 years in atomic physics and quantum optics has lead to the realization of novel quantum phases of matter, including Bose-Einstein condensates~\cite{pita-book}, Fermi degenerate gases~\cite{demarco99,*giorgini2008rmp}, and strongly correlated many-body systems that simulate the behavior of complex models of condensed matter physics~\cite{lewens2012,*bloch2008rmp} and even relativistic quantum mechanics~\cite{gerritsma2011}.
In particular, trapped atomic ions constitute nowadays one of the most advanced platforms for quantum simulation~\cite{blatt2012} and quantum information processing~\cite{haeffner2008}.

In parallel, 
the production of cold molecules has also attracted much attention because of their potential application to quantum information~\cite{DeMille2002} and their sensitivity to the values of fundamental constants~\cite{flambaum07} and parity- and time-violating interactions~\cite{labzo78,*sush78,*kozlov1995,*demille2008}, as well as for the study and control of chemical reactions at ultralow temperatures\cite{krems2008,*bell09b}. In this context, we note the recent improvement in the measurement of the electron's electric dipole moment with a beam of YbF molecules~\cite{hudson2011nat}.
Experimental progress has been steady toward production by a broad range of methods, from photoassociation~\cite{ulmanis2012}, to magnetic-field sweeps through Feshbach resonances~\cite{koehler2006rmp}, buffer-gas cooling~\cite{hutzler2012}, and deceleration of molecular beams by Stark and Zeeman interactions~\cite{vdMeer2012}. 
Of particular interest are cold {\em polar} molecules because of their relatively easy manipulation with external electric fields~\cite{DeMille2002} and because the anisotropic and long-range character of the dipole-dipole interaction makes these systems fundamentally different from cold atomic gases. 

Theoretical studies predict that polar molecules can feature strongly correlated crystalline states and superfluid-crystalline phase transitions~\cite{barnett2006,*astrak2007,*buechler2007,*pupillo2009}. They have also been suggested to simulate quantum magnetism Hamiltonians such as the $XXZ$ and $t$-$J$ models~\cite{gorshkov2011,*maik2012}.
Here, it is worth noting the recent realization of a spin model with KRb molecules in optical lattices, with rotational states playing the role of spins~\cite{yan2013}.

In addition, cold polar molecules have been proposed to realize quantum information tasks either on their own~\cite{DeMille2002} or in hybrid systems with neutral atoms. Particular attention has been paid to the possibility 
of profiting from the strong electric dipole-dipole interaction between polar molecules~\cite{charron2007,*mishima2009} or in hybrid molecule--Rydberg-atom setups~\cite{kuznet2010}. The proposal to use polar molecules together with mesoscopic quantum circuits~\cite{andre2006} has also opened an interesting alternative route toward quantum information processing with molecular species.

For these applications, it is of paramount importance to have an accurate knowledge and control of the properties of the molecules, most notably their electric dipole moments (EDMs). From the theoretical point of view, the determination of accurate EDMs and molecular polarizabilities requires complex calculations~\cite{maroulis2006}. Experimentally, the best measurements to date are usually obtained by molecular-beam electric resonance methods~\cite{brown2003book}. 
These methods are well suited to studying molecules in their ground electronic and vibrational states in a molecular beam. However, it would be interesting to have a tool that can also probe the EDMs of molecules that cannot be produced in beams or in excited rovibrational states, as is usually the case for cold molecules created from cold alkali-metal atoms in photo\-association 
and magneto\-association experiments. 


In this work, we propose an EDM measurement protocol for trapped polar molecules. To do so, we put together two demonstrated techniques\textemdash trapped atomic ions as sensitive probes of weak external forces~\cite{biercuk2010,*hempel2013,murpetit2012,kotler2014}, and ultrafast control on the internal state of cold molecules~\cite{meijer2007}\textemdash to design a quantum-sensing protocol for the measurement of molecular EDMs by coupling a trapped molecule to an atomic ion in a hybrid setup.
We show that, by use of pulsed forces on the ion and the molecule, it is possible to engineer a quantum phase gate between the two.
Interference measurements of this phase on the state of the ion allow then determination of the molecular EDM with an uncertainty similar to that of optical Fourier transform spectroscopy. In addition, this protocol may find further applications in molecular cooling and the quantum simulation of strongly interacting systems of dipoles~\cite{murpetit2014apb}.

In Sec.~\ref{sec:prelim} we present our theoretical approach to describing the hybrid system and the separation of time scales that allows the assignment of a well-defined, nonzero EDM to a trapped molecule.
Based on this framework, in Sec.~\ref{sec:EDM_measurement} we discuss various possible protocols to measure the molecular EDM, including the main result of this paper: a protocol based on using a nearby ion as a quantum probe (Sec.~\ref{ssec:qls_protocol}). To support our claims, we provide numerical results (Sec.~\ref{sec:numerics}) and a detailed experimental proposal (Sec.~\ref{sec:expt_proposal}). Particular elements of the proposal are discussed in further detail in several Appendixes included at the end of the paper.

\begin{figure}[t]
  \centering
  \includegraphics[width=\columnwidth]{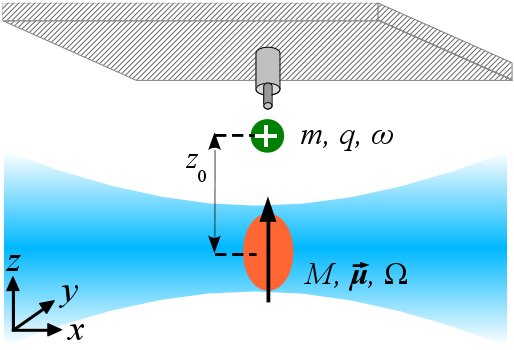}
  \caption{(color online)
    \textit{Scheme of the system.}
    An ion of mass $m$ and charge $q$ (green dot with `+') is confined in an ion trap with a ``stylus electrode'' geometry~\cite{maiwald2009} (gray cylinder) while a polar molecule (orange oval) of mass $M$ and EDM \dipvec\ (thick arrow) is trapped by a focused laser (blue shaded area) a distance $z_0$ below. $\om$ and $\Om$ stand for the corresponding trapping frequencies.}
  \label{fig:system}
\end{figure}


\section{Theoretical approach}\label{sec:prelim}

\subsection{Model of a hybrid ion-dipole system}\label{ssec:model}
We consider an ion of mass $m$ and charge $q$ confined in a harmonic trap of frequency $\om$, and an electric dipole \dipvec\ of mass $M$ in a harmonic trap of frequency $\Om$; see Fig.~\ref{fig:system}. For simplicity, we assume that the traps are spherically symmetric about their minima.
The total energy of the system can be written 
\begin{equation}
  W
  = \half m \om^2 (\bm{x}-\bm{x}_0)^2
  + \half M \Om^2 \bm{X}^2
  + \frac{q\dipvec\cdot
     \left( \bm{x}-\bm{X} \right)}{4\pi\epsilon_0|\bm{x}-\bm{X}|^3}
  \label{eq:energy}
\end{equation}
Here, $\bm{x}$ is the position vector of the ion, $\bm{X}$ the position of the dipole, and $\bm{x}_0=(0,0,z_0)$ the position vector of the ion's trap minimum in the absence of ion-dipole coupling (IDC), while the dipole's trap is taken as the origin of coordinates.

For typical values of the trapping frequencies
 ($\om/2\pi\sim\mathrm{MHz}$, $\Om/2\pi\sim\mathrm{kHz}$),
the IDC will lead to a (small) shift of both particles' equilibrium positions.
The behavior of the system for small displacements around the latter 
can be described in terms of normal modes (NMs)~\cite{murpetit2014apb}. Assuming for simplicity a configuration in which the molecule's EDM is along the axis joining the two trap minima, which we take as the $z$ axis, $\bm{\dip}=\dipz\bm{e}_z$, the motion along each of the $x,y,z$ directions decouples from the others.
We detail in Appendix~\ref{sec:highfreq} a procedure to ensure such alignment initially; Appendix~\ref{sec:expt_uncertainties} contains a discussion of the effects of a possible misalignment.
We then have two NMs of the ion-dipole system in each direction. We name these center-of-mass (\com) and stretch (\str) modes by analogy with the NMs of two-ion systems~\cite{james1998,*kielpinski2000}.
Indeed, 
for the case $\om=\Om$ (when relative and \com\ coordinates exactly decouple), the NMs do correspond to in-phase and out-of-phase displacements of the two particles, as with two ions~\cite{murpetit2014apb}.
It is useful to introduce a parameter $\alpha=q \dip/(4 \pi \hbar \epsilon_0 z_0^2)$, which has units of frequency and gives a measure of the IDC strength.
For typical values (see Table~\ref{tab:numbers})
we find $\alpha/\sqrt{\om \Om} \sim 0.1-10$, which means that one can reach a regime of strong coupling. We also define a characteristic length
$L=[(q\dip/4\pi\epsilon_0)(m\om^2 + M\Om^2)/(mM\om^2\Om^2)]^{1/4} \sim 1-10~\mumeter$.


\subsection{Separation of time scales and effective nonzero molecular EDM}\label{ssec:classical_EDM}

An electric dipole $\dipvec$ in the potential energy landscape given by Eq.~\eqref{eq:energy} and aligned in the direction of the ion has two possible equilibrium positions, $\ztoway$, depending on whether it is pointing toward or away from the ion; see Appendix~\ref{sec:highfreq} and Ref.~\cite{murpetit2014apb}. Assuming that the dipole corresponds to the EDM of a diatomic molecule in a given electronic and vibrational state, its dynamics is governed by the following Hamiltonian
\begin{align}\label{eq:Hmol}
 H_{\mathrm{mol}}
 &=
 \half M \Om^2 \bm{X}^2
 + \sum_{\dip} V_{\mathrm{ext};\dip}(\bm{X}) P_{\dip}
 + \Brot \bm{J}^2  \:.
\end{align}
Here $P_{\dip}$ is the projector on an eigenstate of $\hat{\dip}$, the molecular EDM that couples to external fields, $V_{\mathrm{ext};\dip}$, such as the electric field of a nearby ion. The last term describes the internal energy of the molecule in rotational state \ket{J}, $\bm{J}^2\ket{J} = J(J+1)\ket{J}$.

The EDM of a diatomic molecule in a single rotational state, \ket{J}, is exactly zero in the absence of fields breaking inversion symmetry.
For a hybrid system as in Fig.~\ref{fig:system}, one cannot apply a dc electric field to break the symmetry and hybridize rotational states into pendular states with a nonzero \dip, as the fields required are typically $\sim\mathrm{kV/cm}$~\cite{friedrich99} and might expel the ion from its trap.
Instead, we consider putting the molecule in a superposition of rotational states.
Coherent superpositions of rotational states have been realized in experiments with \KRb~\cite{yan2013}, where coherence times $t_{\mathrm{coh}}\sim 10-100$~ms
were observed, sufficient for implementing the quantum protocol that we introduce below. However, a freely evolving superposition of rotational states would quickly result in a vanishing \textit{average} EDM due to the fast internal dynamics set by the $\Brot\bm{J}^2$ term. To prevent this and keep the EDM oriented with respect to the ion, we propose instead to use two-photon Raman processes with ultrafast pulses.
In the following, we briefly discuss the separation of time scales in $H_{\mathrm{mol}}$ that allows doing so, leaving a more detailed analysis to Appendix~\ref{sec:ultrafast_higherstates}.

The Hamiltonian~\eqref{eq:Hmol} can be split into two parts with very different time or energy scales.
The trapping potential for a given $\dip$, $H_{\dip}^{(0)} = (1/2)M \Om^2 \bm{X}^2 + V_{\mathrm{ext};\dip}(\bm{X})$, has a characteristic energy in the range of
$\Om\mathrm{-}\om \sim \mathrm{kHz-MHz}$~\cite{murpetit2014apb}.
This is much slower than the part of the Hamiltonian describing the internal dynamics, $H^{(1)} = \Brot\bm{J}^2$, characterized by $\Brot \sim \mathrm{GHz}$.
Experiments with molecular beams have proven the possibility of using ultrafast laser pulses, 
of duration $\tau \ll 1/\Brot$, to control the transfer of population between rotational states; see, e.g., \cite{meijer2007}.
We propose to use similar control techniques to effectively ``freeze'' the internal dynamics of the molecule (see the details in Appendix~\ref{sec:bangbang}).
This ultrafast process occurs on a time scale much shorter than the spatial dynamics due to $H_{\dip}^{(0)}$, $\tau \ll 1/\om$. As a consequence, the spatial dynamics of the resulting ``frozen'' dipole, $\dip$, is disentangled from the internal dynamics and governed by $H_{\dip}^{(0)}$.
The coupling of this trapped and oriented EDM with the nearby ion can then be described using the language of normal modes in Sec.~\ref{ssec:model}; see Appendix~\ref{sec:highfreq} and Ref.~\cite{murpetit2014apb}.
Below, we show how, once the molecule is initialized in a state with a nonzero $\dip$, it is then possible to measure the weak electric field generated by this EDM using a trapped ion as a sensitive quantum probe, effectively determining the value of \dip.


\section{EDM measurement protocols}\label{sec:EDM_measurement}
\subsection{Based on existing trapped-ion protocols}\label{ssec:available_prots}
In the absence of an ion, a direct way to measure the EDM of a trapped molecule is to apply a dc electric field $\vec{E}_{\mathrm{dc}}$ on its dipole, and measure the corresponding Stark energy shifts spectroscopically, taking advantage of the long interrogation times in the trap. With both ion and dipole trapped,
one can in principle detect a similar shift due to the ion's field at the dipole's position and, by probing it at various (unknown) distances $z_0$, estimate $\dip$. However, the relative weakness of this effect (see Table~\ref{tab:numbers}) leads us to consider two alternatives that rely on the capability of trapped ions to detect very weak forces~\cite{biercuk2010,*hempel2013,murpetit2012,kotler2014}. 

A first strategy is to characterize the NM frequencies of the coupled system. These depend on the particles' masses, the ratio between trap frequencies $\Om/\om$ and, most fundamentally, on the ratio between IDC and trap energies, $\hbar\alpha/(m\om^2z_0^2) \propto \dip/z_0^4$~\cite{murpetit2014apb}.
By electrically or optically driving the ion, one can excite NMs and determine the resonance frequencies $\omcom$ and $\omstr$ from the oscillations of the ion, thus allowing one to fit
both $z_0$ and $\dip$ as is done in mass spectroscopy in two-ion systems~\cite{drewsen2004,goeders2013}.
These measurements can be done (i) by estimating the heating rate of the ion as a function of the driving frequency~\cite{clark2010}, (ii) by measuring the growth in the number of phonons as a function of the frequency~\cite{haeffner2008}, or (iii) by carefully measuring the ion position and obtaining its Fourier transform.
\begin{table}[tb]
  \caption{Comparison between a system composed of two ions and an ion-dipole system.  Particle no.\ 1 is an ion while no.\ 2 is either an ion or a dipole with $\dip=1$~D, respectively.}
  \label{tab:numbers}
  \begin{tabular}{lcc}
     System & Two ions & Ion + dipole \\ 
  \hline
  Interparticle distance $z_0$
     & $\sim 10\mumeter$ & $\sim 10\mumeter$ \\ 
  Energy of no.\ 1 in field of no.\ 2~~
     & $h\times$35 GHz & $h\times$73 kHz \\ 
  Force of no.\ 1 on no.\ 2
       & 2.3 aN & $4.8\times 10^{-6}$ aN \\ 
  \end{tabular}
\end{table}


\subsection{Quantum-sensing protocol and applications}\label{ssec:qls_protocol}

While the approach based on an analysis of the NM eigenfrequencies is simple, it requires long measurement times ($\sim 0.1-1$~s \cite{drewsen2004,clark2010}) to discern whether a driving is resonant or not, limiting the potential sensitivity to $\dip$.
To overcome this, we propose a method that falls in between the two above: while relying on the NM eigenfrequencies, its goal is to detect the energy shifts and phases induced by the IDC on the trapped ion.
The basic idea is to apply a state-dependent force on the ion causing it to explore two spatial regions where it suffers different energy shifts due to its interaction with the dipole.
Combining this with additional forces on the dipole and initial and final quantum gates on the ion, we can use the internal state of the ion to determine $\dip$. This is similar to previous quantum logic spectroscopy (QLS) protocols for atomic and molecular ions~\cite{murpetit2012,schmidt05} in that the ion acts as a quantum probe of the combined system dynamics.
In those works, the goal was to measure the coupling of a particle to external forces.
For example, Ref.~\cite{schmidt05} relied on the Coulomb force between Be$^+$ and Al$^+$ ions, together with state-dependent forces on both of them, to measure the sensitivity of Al$^+$ to light of a particular frequency.
Instead, here we use the trapped ion as a sensitive measurement device for the weak force that the dipole exerts on it.
In this sense, our proposal is similar to~\cite{kotler2014} in harnessing tools of quantum information science to measure very small energy shifts.
For comparison, while the authors of Ref.~\cite{kotler2014} measured an energy of the order of mHz between two spins separated by $\sim 2$--$3\mu$m, here the aim is to measure an energy shift $\sim$kHz due to the dipole located at a distance $\sim10\mu$m; see Table~\ref{tab:numbers}.

Specifically, our measurement protocol consists of the following steps:
(i) prepare the ion in internal state $\ket{0}_{\mathrm{i}}$;
(ii) apply a $\pi/2$ pulse on the ion, $H=\exp(-i\sigma_{\mathrm{i}}^z \pi/2)$; evolution thereafter of each of its internal states, $\{\ket{0}_{\mathrm{i}},\ket{1}_{\mathrm{i}}\}$, corresponds to an arm in a Ramsey interferometer;
(iii) optionally, apply a reference phase on the ion to maximize the detection signal, $\exp(i\vartheta\sigma_{\mathrm{i}}^z)$;
(iv) apply forces $f_{\mathrm{i,d}}(t)$ on the ion and dipole until the motional state is restored~\cite{leibfried03,garcia-ripoll03,*garcia-ripoll05} and the system experiences a total phase $\exp(i\phi(\dip)\sigma_{\mathrm{i}}^z/2)$ which depends on the molecule's EDM due to the ion-dipole coupling;
(v) finally, close the two arms of the interferometer with a new $\pi/2$ pulse on the ion, and measure the ion internal state.
In contrast to previous QLS works~\cite{leibfried03,garcia-ripoll03,*garcia-ripoll05,murpetit2012}, in this protocol we need only a state-dependent force on the ion, $f_{\mathrm{i}}(t)\sigma_{\mathrm{i}}^z$, while the force on the dipole, $f_{\mathrm{d}}(t)$, may have any origin and does not need to be state dependent.
For simplicity, and to get practical estimates,
below we assume that both $f_{\mathrm{i,d}}$ are optical forces originating from the application of short, far-detuned laser pulses (ac Stark shifts)
on an internal transition of the ion and the dipole, respectively
(in general, the laser systems used to address the ion and the molecule will be different).

The description of the system in terms of NMs allows us to analytically and numerically compute the phase $\phi$ accumulated 
as a function of force duration $T$
and average strength $\overline{f}$~\cite{murpetit2012}:
\begin{equation}
  \phi = \sum_{n=\mathrm{com,str}}\beta_n
    a_n^2 T^2 \overline{f}_{\mathrm{i}}\overline{f}_{\mathrm{d}}
    \,,
    \quad
    a_n^2:=\hbar/\left( m_n \omega_n \right)  
  \label{eq:phase}
\end{equation}
with dimensionless constants $\beta_n\sim O(1)$. 
We remark that the NM frequencies are analytic functions~\cite{murpetit2014apb} of the system $\mycalS=(m,M,\dip;\om,\Om,z_0)$ and driving $\mycalD=(f_{\mathrm{i}},f_{\mathrm{d}},T)$ parameters, i.e.,
$\phi$ is a function $\phi=\phi(\mycalS,\mycalD)$.
We also emphasize that this description is based on the separation of spatial and internal dynamics for the dipole discussed in Sec.~\ref{ssec:classical_EDM}, which imposes the condition that the force duration be much longer than the internal dynamics time scale, $T \gg 1/\Brot$.

This framework allows us to devise several measurement and control applications:
\begin{itemize}
\item[(A)] For periodic drivings, $f_{\mathrm{i,d}} = \overline{f}_{\mathrm{i,d}} \cos(\nu t) e^{-(t/T)^2}$, the accumulated phase diverges for $\nu\approx\omcom,\omstr$, providing an alternative to mode spectroscopy to determine $\omcomstr$.
\item[(B)] If \dip\ is unknown, a measurement of $\phi$ for a given $z_0$ provides an estimate for $\alpha$ and hence $\dip/z_0^2$. The precision with which \dip\ can be determined is then mainly limited by the accuracy in $z_0$.
\item[(C)] More generally, measuring $\phi$ for a range of (unknown) distances, and using the known dependence of $\omcomstr$ on $\alpha$, a multivariate analysis of $\phi$ leads to estimates of \dip\ and $z_0$. 
\item[(D)] Conversely to (B), for systems where \dip\ is known, the protocol allows estimatation of $z_0$, realizing a sort of ``ion-dipole force microscopy'' (IDFM).

\item[(E)] If all system parameters $\mycalS$ are known, one can realize a controlled-phase gate between ion and molecule by properly choosing $T$ and $f_{\mathrm{i,d}}(t)$, using state-dependent forces also on the dipole.
\end{itemize}


\section{Numerical results}\label{sec:numerics}
\subsection{Estimation of molecular EDMs}\label{ssec:molec_params}
We performed numerical simulations to evaluate the feasibility of these applications with two representative model systems, composed of a Ca$^+$ ion and either a KRb or a \CaH\ molecule, thus covering a broad range of EDM values and currently available cold polar molecules.
These calculations 
confirm that one can induce a state-dependent phase $\phi\sim1$~rad on the composite system, similar to what has been realized with atomic ions, thus enabling 
the applications above.

\begin{figure}
  \centering{%
    \includegraphics[width=0.95\columnwidth]{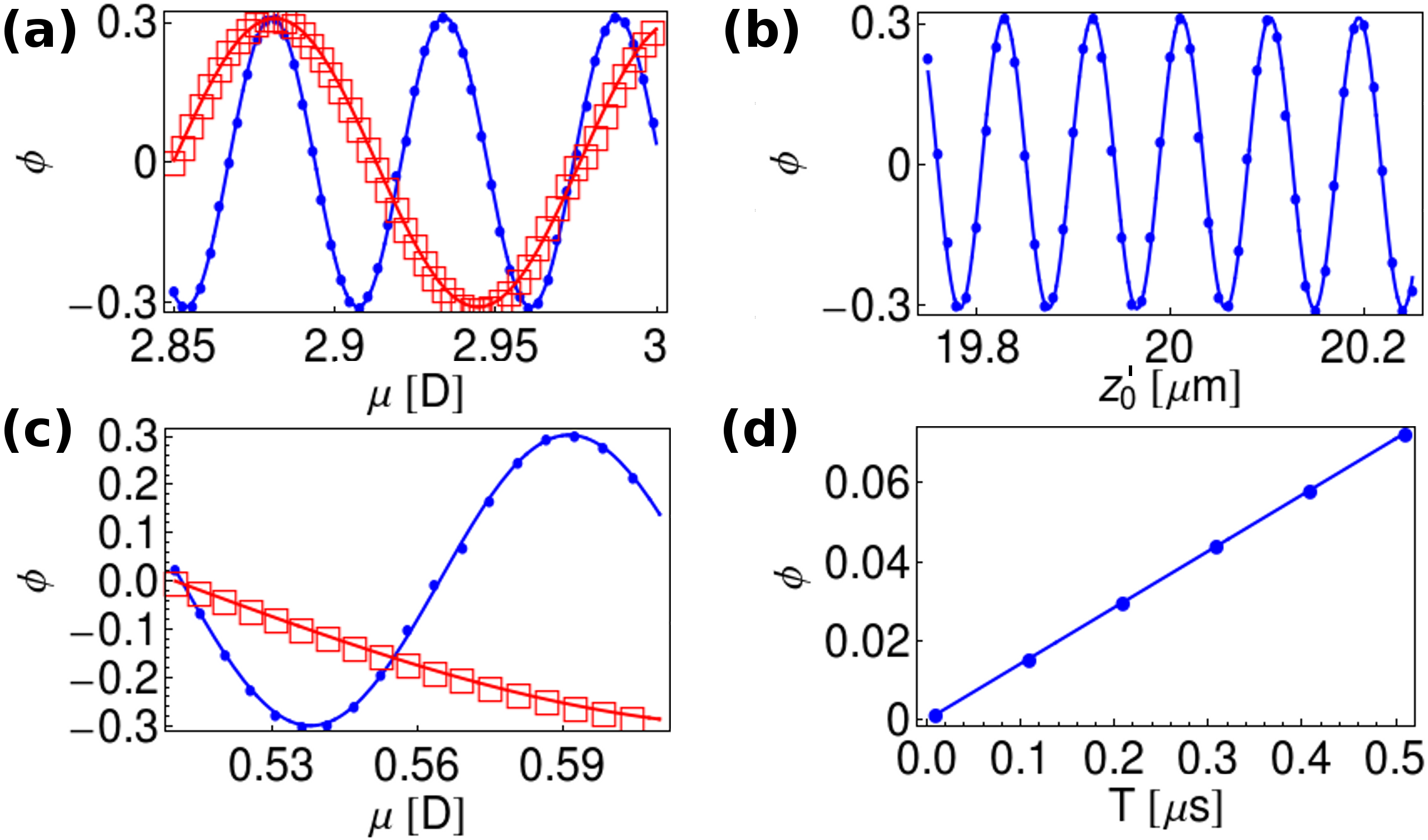}
  }
  \caption{
    (color online) 
     {\bf (a)} Accumulated phase for a \Ca + \CaHgs\ system with $\om/2\pi=1$~MHz,
       $\Om/2\pi=1$~kHz, $T=150$~ns, $\Omega_{\mathrm{Rabi}}/2\pi=300$~MHz,
       $\Delta_{\mathrm{Ca}}/2\pi=0.8$~GHz, and $\Delta_{\mathrm{CaH}}/2\pi=330$~MHz.
       Blue filled circles are calculated for $z_0'=20~\mumeter$, red squares for
       $z_0'=30~\mumeter$; solid lines are sinusoidal fits.
     {\bf (b)} As (a), with $\dip=\dip_{\mathrm{CaH}}$, as a function of
       ion-dipole distance.
     {\bf (c)} The same for \Ca + \KRbgs\ with $T=100$~ns,
       $\Om_{\mathrm{Rabi}}/2\pi=500$~MHz, $\Delta_{\mathrm{Ca}}/2\pi=1$~GHz,
       $\Delta_{\mathrm{KRb}}/2\pi=250$~MHz, and traps as in (a).
     {\bf (d)} As (c) as a function of pulse duration with
       $\dip=\dip_{\mathrm{KRb}}$ and $z_0'=20~\mumeter$.
     Here, $z_0'$ is the actual ion-dipole distance including the displacement due
     to the IDC discussed in Appendix~\ref{sec:highfreq}.
  }
  \label{fig:results}
\end{figure}

Trapping of cold calcium monohydride (CaH) molecules in their ground state was first described in Ref.~\cite{weinstein98}. In their electronic ground state, $X ^2\Sigma^+$, their EDM is $\dip_{\mathrm{CaH}}=2.94$~D~\cite{steimle2004}. 
Rovibrational states within $X ^2\Sigma^+$ should be stable for typical trapping times~\cite{myestim} while radiative lifetimes of the lowest rovibrational levels of the electronically excited state $B ^2\Sigma^+$ are $\tau_\mathrm{rad} \approx 58$~ns~\cite{berg1996}, making them good candidates to implement optical forces.
We plot in Fig.~\ref{fig:results}(a) the phase accumulated, according to Eq.~\eqref{eq:phase}, due to the action of a pair of pulses detuned from Ca$^+$ and \CaH\ resonances, 
for a range of EDMs close to $\dip_{\mathrm{CaH}}$, corresponding to application (B). Note how a small $1\%$ change in \dip\ from its nominal value leads to $\phi$ changing sign and increasing in magnitude. 
On the other hand, Fig.~\ref{fig:results}(b) shows the accumulated phase as a function of ion-dipole distance for $\dip=\dip_{\mathrm{CaH}}$: the sinusoidal fit through the data shows that distances can be retrieved with submicrometer resolution from measurements of $\phi$ [application (D), IDFM].

We consider next a hybrid system composed of Ca$^+$ and KRb molecules.
%
Fermionic $^{40}$K$^{87}$Rb molecules have been produced and confined in harmonic traps~\cite{zirbel2008} as well as optical lattices~\cite{ospelkaus2006}. The EDM of their absolute rovibronic ground state ($X ^1\Sigma^+, v=0, J=0$) was measured as $\dip_{\mathrm{KRb}}=0.566$~D~\cite{ni2008sci}.
%
We show in Fig.~\ref{fig:results}(c) the phase accumulated by a hybrid \Ca + \KRbgs\ system as a function of the molecule's EDM.
We see that for similar trapping parameters as in the Ca$^+$ + CaH case, 
the dependence of $\phi(\dip)$ is smoother, 
due to the smaller value of $\dip_{\mathrm{KRb}}$.
However, similar values for the phase can be reached by an appropriate choice of the excited state to implement the light force.
For example $2(0^+)$ 
($\tau_{\mathrm{rad}} \approx 27$~ns~\cite{ospelkaus2008})
or $(3)^1\Sigma^+$ 
($\tau_{\mathrm{rad}} \approx 0.3$~ns~\cite{aikawa2009}) 
should permit higher intensities resulting in larger phases.

\section{An experimental proposal}\label{sec:expt_proposal}

According to our simulations, the main limiting factor in realizing our protocol is photon scattering by the excited state.
Experience with atomic hyperfine qubits has shown that inelastic off-resonant light scattering can be notably reduced using Raman schemes with large detunings, at the expense of larger laser intensities~\cite{cline1994,*ozeri2005,*ozeri2007};
similar coherent techniques have been applied to ultracold molecules for quantum state transfer~\cite{meijer2007,winkler2007,ospelkaus2008}.
Thus, off-resonant laser pulses appear especially suitable as far-detuned optical forces allow precise bounding of photon scattering probabilities. 
We note that we assumed for simplicity forces that do not depend on the dipole's EDM. However, our formalism can be easily generalized to the case $f_{\mathrm{d}}=f_{\mathrm{d}}(\dip)$.
The main experimental challenges remaining are building a hybrid setup and orienting the molecular EDM for times $\gtrsim T$.
We discuss here an experimental scheme addressing these issues.

We envision the following experimental setup and sequence, cf.\ Fig.\ \ref{fig:system}.
The ion would be confined using a microchip trap~\cite{seidelin2006,person2006,harlander2011} or pure optical means~\cite{schneider2010}.
A particularly attractive setup would use a radio-frequency ``stylus trap''~\cite{maiwald2009} because of its compact design, which allows high optical and spatial access, and high sensitivity to nearby fields.
Below the ion trap, 
a tightly focused laser beam or an optical lattice~\cite{yan2013,deiss2014} would trap the molecule(s).
Additional lasers required for the EDM orientation, rotational state manipulation, and application of optical forces would be directed onto the molecules using a similar optical path.

\begin{figure}
  \centering{%
    \includegraphics[width=\columnwidth]{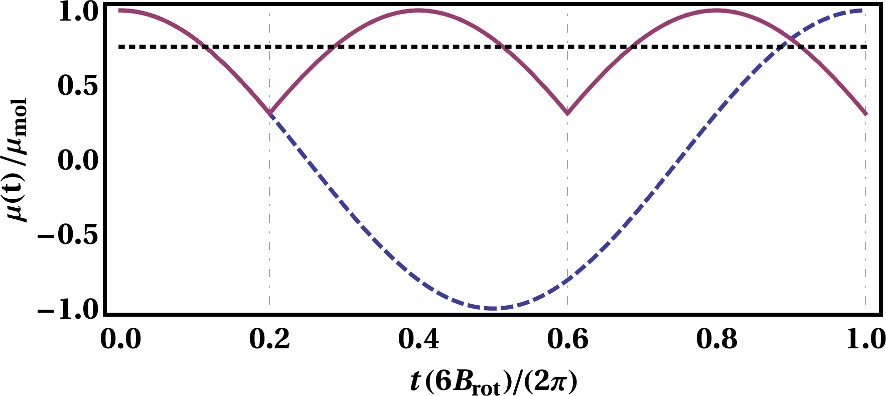}
  }
  \caption{(color online) 
  Instantaneous EDM $\dip(t)$ for a free molecule (dashed blue line)
  and for a molecule subject to $\pi$ pulses at times
  $t=\{1/5,3/5,1\}\times (2\pi/6\Brot)$ (solid red).
  The horizontal dotted line indicates the average value $\overline{\dip}\approx0.76\dip$.
  }
  \label{fig:spinecho}
\end{figure}

To start, as in Ref.~\cite{yan2013}, molecules would be trapped and cooled in their ground electronic, vibrational, and rotational state, $\ground = |X, v=0, J=0\rangle$.
Experiments with cold molecular beams have already demonstrated the possibility of controlling the transfer of population to selected rotational states by means of ultrafast two-photon Raman processes~\cite{meijer2007}, subject to the selection rule $\Delta J=0,\pm2$~\footnote{An alternative scheme to trap and orient molecules in distinct rotational states based on their different polarizabilities was recently reported in Ref.~\cite{deiss2014}.}. Here, we would use the same technique to implement a two-photon $\pi/2$ pulse to transfer the molecules to the state
$|\chi(t=0)\rangle=(\excit+\ground)/\sqrt{2}$, 
with a nonzero \dipvec\ along the $z$ axis; this would complete the initialization of the molecular state.
(Here, $\excit = |X,v=0,J=2\rangle$ is a rotationally excited state of the ground rovibronic manifold.)
As noted above, in free space, the energy difference between \ground\ and \excit\ would lead to fast oscillations of $\dip(t)=\langle\chi(t)|\hat{\dip}|\chi(t)\rangle$, which would quickly average out.
To prevent this, one would ``freeze'' the molecule's internal dynamics by applying a train of $\pi$ pulses at a rate $\nu_\pi > 6B_{\mathrm{rot}}$ (for details, see Appendix~\ref{sec:bangbang}), which amounts to a dynamical decoupling scheme~\cite{NMRbook,viola1998}. Such ultrafast manipulation strategies have been implemented with sequences of microwave~\cite{biercuk2009,szwer2011} or optical~\cite{tan2013,hayes2012,hucul2015} pulses in trapped-ion experiments, and could be realized with the same Raman lasers used for the state initialization. 
As shown in Fig.~\ref{fig:spinecho}, this strategy results in a nonzero average EDM [cf. Eq.~\eqref{eq:ave_dip}],
\begin{align}
 \overline{\dip}=6\Brot \int_{0}^{1/(6\Brot)} \dip(t) dt
 \approx \dipmol[1 + O(\Brot/\nu_\pi)^2] \:,
 \label{eq:ave_dip_mainbody}
\end{align}
on which the much slower off-resonant pulses required for the phase gate can be applied.
Indeed, the very long natural lifetimes of rotational states in the lowest vibrational manifolds of the ground electronic state of polar molecules~\cite{myestim}, together with their observed coherence times $t_{\mathrm{coh}} \approx 10$--$100$~ms~\cite{yan2013}, open a wide pulse-duration window, $1/\Brot \ll T \ll t_{\mathrm{coh}}$, 
to manipulate and measure the molecule's EDM.
Within this frame, the aforementioned electronic excited states appear suitable to implement the light forces required by our quantum-sensing protocol.
Taking these constraints into account, we calculated the phase accumulated by a Ca$^+$ + KRb system as a function of pulse time $T$ [application (E)]. As the results in Fig.~\ref{fig:results}(d) show, phases of order 0.1~rad can be generated.
Interestingly, the fact that the present protocols enable pulses much shorter than the
oscillation period due to dipole-dipole interactions between nearby molecules, $t_{\mathrm{dip}} \approx 20$~ms~\cite{yan2013}, opens the perspective to study the coherent dynamics of molecules \textit{in real time}~\cite{rabitz2010,*bustard2011,shapiro2012book}.
This can be of interest for nonequilibrium quantum simulations with polar molecules~\cite{hazzard2013}.

\section{Summary and outlook}\label{sec:conclusions}

In summary, we have studied a hybrid quantum system composed of an ion and an electric dipole, and shown how the ion can be used as a sensitive probe of the dipole's magnitude. To this end, we have relied on the use of ultrafast control pulses to effectively orient the dipole, and then proposed a quantum protocol to retrieve the information on the dipole's electric moment encoded in the ion's state.
We have provided numerical calculations demonstrating the feasibility of a range of applications, from measuring EDMs to mapping the distribution of dipoles, with experimental tools currently available (separately) in trapped-ion and cold-molecule laboratories.
Moreover, we have seen that a regime of strong ion-molecule coupling can be achieved.
This may allow for further molecule cooling methods by interaction with trapped ions, as well as the creation of ion-molecule entangled states~\cite{murpetit2014apb,huber2012}.

The uncertainty in the EDM value achievable with this protocol is proportional to the uncertainty $\delta \phi$ in the estimation of the accumulated phase~\cite{murpetit2014apb}. Setting aside quantum-metrology schemes requiring probe entanglement, this is limited by quantum projection noise on ion state detection~\cite{itano1993-qpl}, according to which $\delta \phi \propto 1/\sqrt{N}$, with $N$ the number of measurements.
Molecular EDMs measured by optical Stark spectroscopy of diatomic molecules typically are obtained with an uncertainty on the order of $10^{-2}-10^{-3}$ (see, e.g., \cite{wang2010a,*wang2010b}), a precision which should be achievable with the present protocol.
On the other hand, determinations of EDMs of larger molecules by molecular beam Fourier transform spectroscopy in an electric field in the microwave range reach nowadays relative uncertainties on the order of $10^{-4}$ (see, e.g., \cite{wohlfart2008,*filsinger2008,*krasnicki2010}). This method requires fitting the observed data to the molecular Hamiltonian including the Stark effect, and knowledge of molecular-structure parameters with sufficient accuracy; its precision is ultimately limited by electric-field inhomogeneities and the interrogation time of the molecules as they fly through the detector.
It appears difficult to reach a similar precision with the protocol proposed here with current coherence times, which limit the number of measurements. On the other hand, the present scheme does not require an accurate prior knowledge of the molecule's structure parameters and model Hamiltonian. Furthermore, it enables the study of molecular species that cannot be produced as a molecular beam, as is the case with ultracold heteronuclear dimers.

We anticipate that these tools will enable additional realizations of quantum information tasks with hybrid systems~\cite{wallquist2009}, in particular for the quantum simulation of quantum mag\-netism and far from equilibrium dynamics~\cite{hazzard2013}, e.g., relying on novel surface traps able to conduct microwaves and realize large magnetic-field gradients~\cite{ospelkaus08,*warring2013,*belmechri2013}.
Finally, the protocols discussed here also boost the possibilities of molecular coherent control~\cite{shapiro2012book,rabitz2010,*bustard2011} and coherent conversion of radiation between optical and microwave frequencies~\cite{barna2012}.

\begin{acknowledgments}
We acknowledge useful discussions with J. Deiglmayr, J. Ortigoso, and M. Schnell.
This work was supported by Spanish MINECO Project No. FIS2012-33022,
ESF COST Action IOTA (MP1001),
US National Science Foundation (Grant No.\ NSF PHY11-25915),
and the JAE-Doc Program (CSIC and European Social Fund). 
JMP acknowledges useful discussions with the participants of the KITP Programs ``Fundamental Science and Applications of Ultracold Polar Molecules'' and ``Control of Complex Quantum Systems,'' and KITP staff for help and hospitality during his stay.
\end{acknowledgments}

\appendix


\section{Effective EDM for a trapped dipole}
\label{sec:highfreq}

In the absence of external fields that break symmetry under parity, a diatomic molecule in a pure vibrational-rotational state $|v,J\rangle$ has no permanent EDM for symmetry reasons, $\langle v,J|\hat{\dip}|v,J\rangle = 0$.
To induce an EDM in such a molecule, one can transfer it into a superposition state, such as
\begin{align}
  |\psi\rangle = \left( |0,0\rangle + |0,2\rangle \right)/\sqrt{2} \,.
  \label{eq:superposition}
\end{align}
In this state, indicating the transition dipole moment by $\dipmol=\langle0,0|\hat{\dip}|0,2\rangle \in\mathbb{R}$, the molecule will have an EDM given by
\begin{align}
  \dip = \langle\psi|\hat{\dip}|\psi\rangle
  = \dipmol \neq 0 \,.
  \label{eq:dip_static}
\end{align}
The two rotational states have different energies, $\Brot \bm{J}^2\ket{v,J} = \Brot J(J+1)\ket{v,J}$. Hence, as time goes by, the two rotational components in~\eqref{eq:dip_static} will acquire different phases, which results in an oscillating EDM of the form
\begin{align}
  \dip(t) &= \dipmol\cos(\erot t/\hbar)  \,,
  \label{eq:dip_dyn}
\end{align}
with $\dipmol$ given by Eq.~\eqref{eq:dip_static} and
$\erot=E(\ket{0,2})-E(\ket{0,0})=h\times 6\Brot$, where $h$ is Planck's constant.
Diatomic molecules typically have rotational constants of a few to a few hundreds of gigahertz (e.g., $\Brot(\mathrm{KRb})=1.114$~GHz~\cite{ni2008sci}, $\Brot(\mathrm{CaH})=128$~GHz~\cite{steimle2004}).
Thus, the dynamics of the system as given by $H_{\mathrm{mol}}$, Eq.~\eqref{eq:Hmol}, is characterized by a series of motions at very different frequencies: $\Om \sim \mathrm{kHz} \ll \om \sim \mathrm{MHz} \ll \Brot \sim \mathrm{GHz}$.

Due to the sinusoidal dependence in Eq.~\eqref{eq:dip_dyn}, in the absence of external trapping, the electric field generated by the dipole on the ion would average to zero over a rotational period $h/\erot$.
The presence of the trapping potential, however, renders inequivalent the situations when the dipole is pointing toward the ion ($\dipvec\cdot\rvec = \dip z$), 
and when it is pointing away from it  ($\dipvec\cdot\rvec = -\dip z$).

To see this, let us look at the total energy of the interacting ion-dipole system, written in terms of relative, $\rvec=\x-\X$, and center of mass (\com), $\R=(m\x + M\X)/(m+M)$, coordinates (cf.~\cite{murpetit2014apb})
\begin{align}
  W
  &=
   \half m \om^2 (\R-\bm{x}_0)^2
  + \half M \Om^2 \R^2
  \nonumber \\
  &+ \half \mred \omstrz^2 \rvec^2
  - \mred \om^2 \z_0\cdot\rvec
  + \frac{q\dipvec\cdot\rvec}{4\pi\epsilon_0|\rvec|^3}
  \nonumber \\
  &+ \mred(\om^2-\Om^2) \R\cdot\rvec \:,
 \label{eq:energyrR}
\end{align}
where we introduced $\mtot=m+M$ as the total mass
and the reduced mass $\mred=m M/\mtot$,
and we identified the relative-motion collective mode (``stretch mode'') frequency for the uncoupled ($\dip=0$) and overlapping ($\z_0=0$) system: $\omstrz^2 := (m\Om^2 + M\om^2)/\mtot$.
The terms on the second line of Eq.~\eqref{eq:energyrR} correspond to the relative coordinate being in a harmonic potential displaced from the origin, while the coupling with \dipvec\ amounts to a further displacement.
For the usual case $\om \gg \Om$, it follows that $\omstrz \sim \om$, i.e., the trapping frequency of the relative motion will be of the same order as $\om \sim \mathrm{MHz}$.
Finally, the last term in~\eqref{eq:energyrR} indicates the coupling between relative and \com\ motions, present if (and only if) the two trapping frequencies differ.

Let us assume for the moment that $\Om=\om$ (hence, $\omstrz=\om$), so that the last term in $W$ [Eq.~\eqref{eq:energyrR}] vanishes.
Then, the IDC has the effect of a potential added on top of the displaced harmonic trap.
Its magnitude can be estimated by introducing
\begin{align}  
 \rdip &:= \frac{q\dip}{4\pi\epsilon_0 z_0^3}\frac{1}{m\om^2}
        = \frac{\hbar\alpha}{m\om^2z_0^2} z_0
        \equiv \talpha z_0 \:,
 \label{eq:rdip}
\end{align}
so that $q\dipvec\cdot\rvec/(4\pi\epsilon_0 |\rvec|^3) \approx \pm m\om^2\rdip z/2$, with the sign depending on the orientation of \dipvec\ with respect to the ion.
As a consequence, the instantaneous potential energy minimum for the dipole is located at $\z_0' = \z_0 \pm \rdip\bm{e}_{\dip}(t)$, where $\bm{e}_{\mathrm{ion}}(t)$ is the unit vector pointing in the direction joining ion and dipole.
This means that, when the dipole points toward the ion, its potential energy minimum is at $\ztow=z_0+\rdip$ and it generates a field $\Vtow=q\dip/(4\pi\epsilon_0\ztow^2)$ on the ion; when it points away from the ion, its equilibrium position is $\zaway = z_0 - \rdip$ and it generates $\Vaway=q\dip/(4\pi\epsilon_0\zaway^2)$.

If we consider a molecule in the superposition state~\eqref{eq:superposition} initially located at \ztow, because of the time dependence of its EDM, it will move about trying to reach \zaway\ half a rotational period later.
Even though the distance between these positions is relatively small, it is still a sizable fraction of the ground-state spread of the molecule in its harmonic trap, $a_{\mathrm{mol}}=\sqrt{\hbar/(M\Om)}$, see Table~\ref{tab:lengths}.
This, together with the high value of $\Brot$, would render the spatial motion of such a dipole nonadiabatic. To avoid this, we rely on ultrafast Raman pulses to ``freeze'' its free evolution, as detailed in Appendix~\ref{sec:bangbang}.

\begin{table}[tb]
  \caption{Typical length scales for the ion--polar-molecule systems studied, for trapping frequencies $\om/2\pi=1$~kHz for the Ca$^+$ ion and $\Om/2\pi=1$~kHz for the molecules (CaH, KRb).}
  \label{tab:lengths}
  \begin{tabular}{c|cc|c|c}
     $a_{\mathrm{ion}}$ &  \multicolumn{2}{c|}{$a_{\mathrm{mol}}$} & $z_0$ & $\rdip$ \\  \hline
                        &  CaH & KRb  & & \\
     16~nm & 496~nm & 282~nm & 10~\mumeter & 10~nm
  \end{tabular}
\end{table}

\section{Dynamical Decoupling of high-frequency EDM oscillations}\label{sec:bangbang}

A key point when measuring a molecular EDM is the difficulty of determining its alignment with respect to a known axis. A strategy to solve this is to apply a strong dc electric field, which polarizes the molecule, resulting in an easy mapping of the molecular frame to the laboratory frame. As discussed in the main text, this strategy is not suitable for a setup with a nearby ion, which would be expelled from its trap.

\begin{figure*}[tbh]
  \centering
  \includegraphics[width=\textwidth]{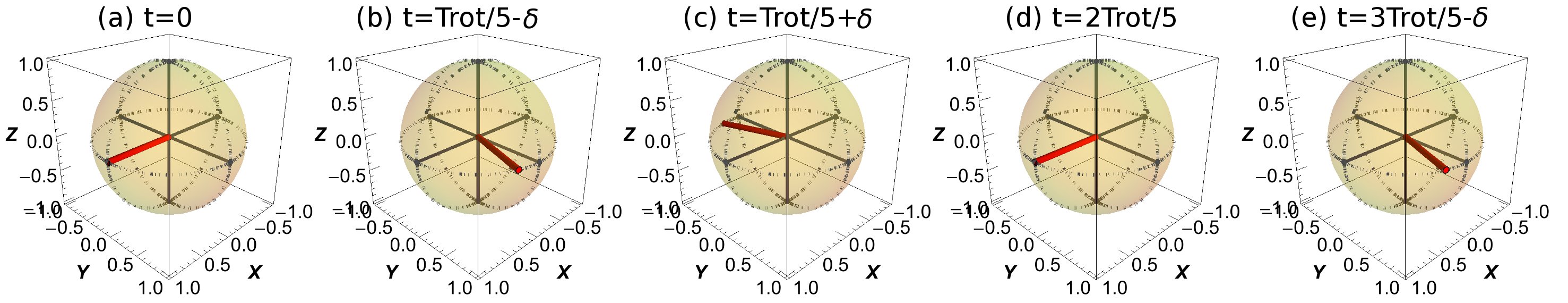}
  \caption{%
  (color online)
    Time evolution of the initial state $\ket{\chi(t=0)}=\ket{+\sigma_x}$ represented on the Bloch sphere (yellow) in spin space, under the sequence of pulses described in Appendix~\ref{sec:bangbang}.
    The position of the state vector is indicated each time by a thick red line.
    \textbf{(a)} Initial state;
    \textbf{(b)} state just before the first $\pi$ pulse at time $t=\Trot/5-\delta$ ($0<\delta\ll\Trot$);
    \textbf{(c)} just after the first $\pi$ pulse at time $t=\Trot/5+\delta$;
    \textbf{(d)} at time $t=2\Trot/5$: it has come back to $\ket{+\sigma_x}$;
    \textbf{(e)} at time $t=3\Trot/5-\delta$, just before the next $\pi$ pulse.
    $\Trot=h/\erot$ is the time for a full rotation in the absence of pulses; cf.\ Eq.~\eqref{eq:dip_dyn}.
  }
  \label{fig:bloch}
\end{figure*}

Instead, we take advantage of the fact that at the very low temperatures of ultracold molecule experiments, one can populate a single rotational state \ket{J} within the vibronic (i.e., vibrational and electronic) ground state.
Then, taking into account the strong anharmonicity of rotational spectra,
$E(J)=\Brot J(J+1)$, one has well-characterized transitions between single rotational states, which can be addressed using microwave, radio frequency, or two-photon stimulated Raman transitions.
We consider for concreteness manipulations between the two rotational states,
$\ket{J=0, M_J=0}\leftrightarrow\ket{J=2,M_J}$, with $M_J$ the projection of the rotational angular momentum on the space-fixed $z$ axis; the exact value of $M_J$ for the second state will be determined by the Raman lasers polarizations.
Because of the long lifetimes of these states, the molecule can be considered as a closed two-level system,
with states $\ket{\JzeroSpin} \equiv \ket{J=0,M_J=0}$,
and $\ket{\JtwoSpin} \equiv \ket{J=2,M_J}$.
Any superposition of these states, $\ket{\chi}=\alpha \ket{\JzeroSpin}+ \beta \ket{\JtwoSpin}$, can be represented on the Bloch sphere in the usual way~\cite{AllenEberly,GerryKnight}.
(The role of other rotational states is analyzed in Appendix~\ref{sec:ultrafast_higherstates}.)

The interaction of the two-level molecule with an intense radiation field can be modeled with the Jaynes-Cummings Hamiltonian in the rotating-wave approximation (RWA)~\cite{AllenEberly,GerryKnight},
\begin{align}
 H           &= (\erot/2) \sigma_z + \hbar\zeta(t) \sigma_x
 \label{eq:hamRWA} \\
 \sigma_Z    &= \ketbra{\JtwoSpin} - \ketbra{\JzeroSpin} \\  
 \sigma_X    &= (\ket{\JtwoSpin}\bra{\JzeroSpin} + \ket{\JzeroSpin}\bra{\JtwoSpin})/2
\end{align}
where we introduced $\hbar\zeta(t)  = \dip_{\mathrm{mol}}{\cal E}(t)$ 
with $\dip_{\mathrm{mol}} = \bra{\JtwoSpin} \hat{\dip} \ket{\JzeroSpin}$ [cf. Eq.~\eqref{eq:dip_static}], and $\cal E$ is the laser electric field.
For our purposes, it is convenient to consider the Raman lasers as linearly polarized along the real-space $z$ direction. Then, the upper rotational state coupled with \ket{J=0} will be \ket{J=2,M_J=0}. We make this choice because of our interest in a particular superposition state,
$
(\ket{J=2,M=0} +\ket{J=0})/\sqrt{2}$,
which has a nonzero EDM pointing along the real-space $z$ axis, i.e., along the direction joining the molecule and the ion.
Its spin-space representation is
$\ket{+\sigma_X}:=(\ket{\JzeroSpin}+\ket{\JtwoSpin})/\sqrt{2}$,
i.e., it corresponds to the point on the Bloch sphere crossing with the `$+X$' axis in spin space; see Fig.~\ref{fig:bloch}(a).

The first term in Eq.~\eqref{eq:hamRWA} reflects the fact that the two rotational states have energies differing by $\erot=h\times6\Brot$ and, in the Bloch picture, generates rotations of the state vector $\ket{\chi}$ around the $Z$ axis (in the fictitious spin-1/2 space!) at a rate $\erot/h$.
The second term, which describes the interaction with the radiation field, corresponds to rotations around the $X$ axis at a rate $\zeta$.

Our strategy to have an average nonzero EDM is then based on ideas of dynamical decoupling~\cite{NMRbook,viola1998} and relies on the following properties of the evolution of the molecular state in the Bloch picture:
\begin{enumerate}
\item Initialize the molecule in state $\ket{\chi(t=0)} = \ket{+\sigma_X}$.
Starting with a molecule in $\ket{\JzeroSpin} = \ket{J=0,M_J=0}$, this is accomplished with a $\pi/2$ pulse that, in spin space, rotates the state vector to \ket{+\sigma_X}; see Fig.~\ref{fig:bloch}(a).
\item Free evolution makes the state vector rotate on the $X$-$Y$ plane toward the $Y$ axis; see Fig.~\ref{fig:bloch}(b).
\item A rotation of $\pi$~rad around the $X$ axis at any point in time, moves the state vector from the position on the $X$-$Y$ plane determined by the polar angle $\varphi$ to $-\varphi$; see Fig.~\ref{fig:bloch}(c).
\item Free evolution from that position is again a rotation around $Z$ in the same direction as before; see Figs.~\ref{fig:bloch}(d) and \ref{fig:bloch}(e).
\end{enumerate}
As the time for a complete rotation on the $X$-$Y$ plane in the absence of such control pulses is $\Trot:=h/\erot=1/(6\Brot)$, it follows that submitting the molecule to a $\pi$ pulse at a rate $\nu_{\pi} > 1/\Trot = 6\Brot$ will result in the average molecular EDM being different from zero.

To see this, consider the spin-space basis given by
$\ket{+\sigma_X},\ket{-\sigma_X}$, with $\ket{-\sigma_X}:=(\ket{\JtwoSpin}-\ket{\JzeroSpin})/\sqrt{2}$.
Noting that the EDMs of these states are
$\dip_{\pm} = \bra{\pm\sigma_X}\hat{\dip}\ket{\pm\sigma_X}=\pm\dip_{\mathrm{mol}}$,
we find that the molecule's EDM at any time is given by
\begin{align}
 \dip(t)
 &= \dip_+|\bra{\JzeroSpin}\chi(t)\rangle|^2 + \dip_-|\bra{\JtwoSpin}\chi(t)\rangle|^2
 \nonumber \\
 &= \dip_{\mathrm{mol}}
   \left(|\bra{\JzeroSpin}\chi(t)\rangle|^2 - |\bra{\JtwoSpin}\chi(t)\rangle|^2 \right)
 \nonumber \\
 &= \dip_{\mathrm{mol}} \left(1 - 2|\bra{\JtwoSpin}\chi(t)\rangle|^2 \right)
\end{align}
Therefore, starting with $\ket{\chi(t=0)}=\ket{\JzeroSpin}$ and preventing the state vector from reaching \ket{\JtwoSpin}, $|\bra{\JtwoSpin}\chi(t)\rangle|<1$ for all $t$, leads to a nonzero average value of $\dip(t)$.
Use now of the sequence of dynamical-decoupling pulses outlined above,
with pulses occurring at time $t_1$ and $t_k=t_{k-1} + 1/\nu_\pi$, 
leads to an average EDM
\begin{align}\label{eq:ave_dip1}
 \overline{\dip}
 =\frac{1}{\Trot} \int_{0}^{\Trot} \mu(t) \, dt
 = \dip_{\mathrm{mol}} \frac{\sin(t_1 \erot/\hbar)}{t_1 \erot/\hbar} \:.
\end{align}
For instance, with $t_1=\Trot/5$ and $\nu_{\pi}=1/(2t_1)=2.5\erot/h$, 
the result is $\overline{\dip}\approx0.76\dip_{\mathrm{mol}}$
; this is the result shown in Fig.~\ref{fig:spinecho} of the main body of the paper.
In the limit of many pulses per rotational cycle, $\nu_{\pi} \gg \Brot$, one gets
\begin{align}\label{eq:ave_dip}
 \overline{\dip}
 &\approx
   \dip_{\mathrm{mol}}
   \left[ 1 - \frac{1}{3!} \left(
         \frac{\Brot}{2\nu_{\pi}} \right)^2 + \cdots \right]  \,,
\end{align}
which is the result quoted in Eq.~\eqref{eq:ave_dip_mainbody} of the main text.

Let us remark that our results are based on a rudimentary pulse sequence, with instantaneous pulses occurring at time $t_1=\Trot/5$ and then at a fixed rate $\nu_\pi=5\Brot/2$, cf.\ Fig.~\ref{fig:bloch}.
It is worth noting that more elaborate pulse sequences have been designed which are more robust against experimental fluctuations in timings, pulse durations, and other sources of error, such as the Carr-Purcell-Meiboom-Gill (CPMG) sequence~\cite{carr1954,*meiboom1958,NMRbook} or the Uhrig Dynamical Decoupling (UDD) scheme~\cite{uhrig2007}, which has been implemented with trapped ions~\cite{biercuk2009,szwer2011} and cold atoms~\cite{yu2013}. More advanced and robust strategies have also been developed (see e.g.~\cite{brown2004,bermudez2012}) and implemented~\cite{tan2013} in the ultracold regime, but their discussion lies beyond the scope of this work.


\section{Ultrafast pulses and role of higher-lying rotational states}
\label{sec:ultrafast_higherstates}

The main difficulty in implementing the dynamical-decoupling protocols in Appendix~\ref{sec:bangbang} for aligning a molecule lies in the high-frequency requirement set by the rotational constant $\Brot \sim \mathrm{GHz}$: one has to couple two states separated by a GHz-energy difference at a GHz rate, i.e., using pulses with duration $\tau \ll 1/\Brot \sim 1~\mathrm{ns}$. 
For CaH (\Brot=128 GHz), one needs $\tau < 3.4$~ps; for KRb (\Brot=1.14 GHz), $\tau < 0.4$~ns.
In Ref.~\cite{santam2011}, rotational transitions in metastable CO
$a^3\Pi_1 |v=0\rangle$ 
trapped on a chip were driven with microwave pulses of duration $\tau \sim 1~\mu$s, yet we are not aware of pulsed microwave sources that fulfill our protocol's time-scale needs.
However, these time-scale and energy requirements can be met using two-photon Raman pulses with ultrafast lasers addressing selected excited rotational-electronic states.
The feasibility of this proposal is based on the results from two independent experimental groups: 
selective population of particular rotational states of cold molecular beams, demonstrated with NO($X ^2\Pi$) molecules by van der Zande and Vrakking and co-workers~\cite{meijer2007}; and the ultrafast quantum-control experiments of atomic hyperfine qubits realized with the aid of a frequency comb by Monroe and co-workers~\cite{campbell2010,mizrahi2014}.
We mention also the large body of research available on coherent control of molecular rotational states; see, e.g., \cite{shir2008,ohshima2010,lemeshko2013} for recent reviews.

Ultrafast laser pulses of duration $\tau \leq 1/\Brot$ have bandwidths $\delta\om \sim 1/\tau \geq \Brot$.
In contrast to the situation encountered with atoms or atomic ions, pulses with such bandwidths can in principle couple a given state to several rotational states, which are spaced by multiples of $\Brot$. One should consider whether it is possible to drive transitions between the states of interest, $\ground,\excit$, without population leaking out to unwanted rotational states. We study in the following a minimal five-level space to assess this question.

\begin{figure}[bt]
  \centering
    \includegraphics[width=\columnwidth]{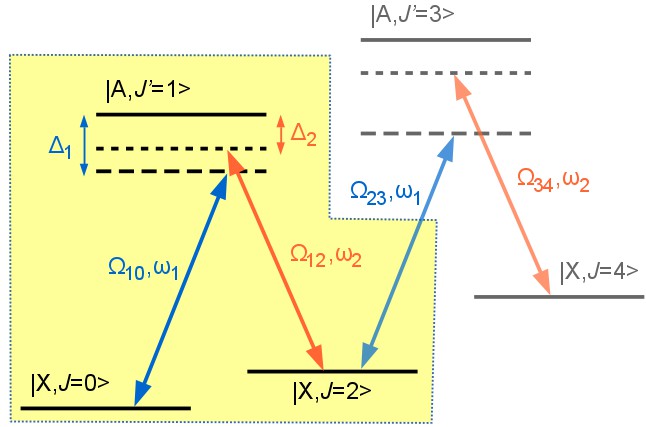}
  \caption{ (color online)
    \textit{Model five-level system}.
    The two rotational states used to construct a state with nonzero EDM are $\ket{X,J=0}$ and \ket{X,J=2}. These states are coupled by two lasers in a Raman scheme (colored arrows) via \ket{A,J'=1}. These three states form the Hilbert space of interest, indicated by the shadowed box.
    In the electric-dipole approximation, state \ket{X,J=2} can leak outside this space to \ket{A,J'=3}, and this to \ket{X,J=4}.
    Relevant laser frequencies, $\om_{1(2)}$, Rabi frequencies, $\Om_{mk}$ [Eq.~\eqref{eq:Om_mk}], and detunings, $\Delta_{1(2)}$, are also identified.
    Note that the nonharmonic character of the rotational spectrum renders the detunings and, hence, Rabi frequencies for transitions to and from \ket{A,J'=3} generally different from those        \ket{A,J'=1}.}
  \label{fig:leakage}
\end{figure}

We consider that our Hilbert space of interest is formed by three states, starting in the lowest rotational state of the ground electronic state of the molecule, which we label $\ket{0}\equiv\ket{X,J=0}$, and which corresponds to state $\ket{\JzeroSpin}$ in Appendix~\ref{sec:bangbang}. Population from this state will be driven by a two-photon Raman transition to state $\ket{2} \equiv \ket{X,J=2} \equiv \ket{\JtwoSpin}$ via an intermediate, electronically excited state $\ket{1}\equiv\ket{A,J'=1}$; here $A$ stands for a generic electronically excited state, 
depending on the molecule and laser employed. Such transitions are allowed in the dipole approximation by the selection rule $\Delta N=0,\pm1$~\cite{bunker1998}; for the case of singlet states, $N=J$.
Now, the same selection rules allow in principle population from \ket{2} to ``leak out'' to $\ket{3}=\ket{A,J'=3}$, and from there to $\ket{4}\equiv\ket{X,J=4}$, and so on, as sketched in Fig.~\ref{fig:leakage}.

Working in the dressed-atom picture with at most one photon being absorbed from/emitted into the $\om_{1,2}$ fields, the relevant Hamiltonian describing this five-level system in the rotating-wave approximation can be written in the form
[in the basis $\{ \ket{0}, \ket{1}, \ket{2}, \ket{3}, \ket{4} \}$]
\begin{align}\label{eq:Ham5LS}
 H
 &=
 \left(
 \begin{array}{ccccc}
    \Delta_1   & \Om_{10}   & 0          & 0                & 0 \\
    \Om_{10}^* & -i\Gamma_A & \Om_{12}   & 0                & 0 \\
    0          & \Om_{12}^* & \Delta_2   & \Om_{32}         & 0 \\
    0          & 0          & \Om_{32}^* & E_3-i\Gamma_A  & \Om_{34} \\
    0          & 0          & 0          & \Om_{34}^*       & E_4
 \end{array} \right) \:.
\end{align}
Here, we have removed a constant term $E_1=2B_A$, where $B_A$ is the rotational constant of the excited state $A$. 
$\Gamma_A$ is the spontaneous decay rate from state $A$, and 
we neglect decay from states within $X$.
$\Om_{mk}$ is the laser-induced dipole coupling between states $\ket{m} \in A$ and $\ket{k} \in X$,
\begin{align}
  \Om_{mk} = \hbar^{-1} \bra{m} \dip\cdot{\cal E} \ket{k} \:,
  \label{eq:Om_mk}
\end{align}
with ${\cal E}$ the laser field at the molecule's position.
Finally, $E_{3(4)}$ is the energy of state \ket{3} (\ket{4}) in the RWA; they are parametrized in terms of the rotational constants of the ground and excited electronic states, $B_{X,A}$, and the photon energies, $\omega_{1,2}$.

\begin{figure}[tb]
  \centering
    \includegraphics[width=\columnwidth]{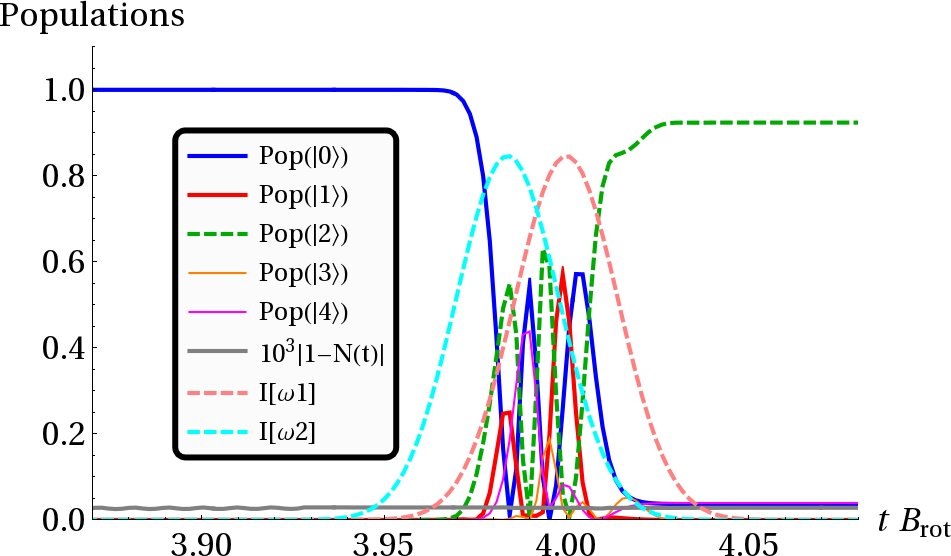}
  \caption{ (color online)
    Numerical evolution of the 5-level system described by Eq.~\eqref{eq:Ham5LS} starting with all population in state \ket{0} (solid blue line) using the parameters in Table~\ref{tab:params5LS}.
    At the end of the simulation, $\approx 92\%$ population is in \ket{2} (dashed green line), while less than $\approx 3.7\%$ has remained in \ket{0} and $\approx 3.9\%$ has leaked to \ket{4} (solid purple line).
    After the pulses, population in states \ket{1,3} (solid red and orange lines) is negligible due to the large value of the detuning.
    The gray line with small wiggles shows $10^3|1-N(t)|$, with $N(t)$ the total population at time $t$.
    The dashed Gaussian profiles centered around $t \approx 4$ stand for the laser intensity profiles (in arbitrary units).
  }
  \label{fig:evol5LS}
\end{figure}

For the dynamical-decoupling protocol in Appendix~\ref{sec:bangbang} to work, we need to be able to drive population between \ket{0} and \ket{2} without having transferred substantial population to states \ket{4}, \ket{1}, or \ket{3} at the end of the pulse.
In Fig.~\ref{fig:evol5LS} we show numerical results corresponding to the time evolution with Hamiltonian~\eqref{eq:Ham5LS} with the parameters indicated in Table~\ref{tab:params5LS}. For this simulation, we have taken all population initially in state \ket{0} and used two delayed Gaussian pulses to drive the transitions $\ket{0} \leftrightarrow \ket{1}$ and $\ket{1} \leftrightarrow \ket{2}$, respectively,
\begin{align}
 {\cal E}_j(t)
    &= {\cal E}_j^{(0)} \exp\{-[(t-t_{j}^{(0)})/\tau]^2\} \cos(\om_j t+\phi_j) \:, \\
 t_{1}^{(0)} &= t_0 \:, \qquad 
 t_{2}^{(0)} = t_0 + \delta_\textsc{stirap} \:.
\end{align}
Note that a value $\delta_\textsc{stirap}<0$ indicates counterintuitive pulse order, i.e., the pulse at frequency $\om_2$ arrives before that at $\om_1$, as in stimulated Raman processes (STIRAPs).
Following~\cite{meijer2007}, for this calculation we have fixed the delay to
$|\delta_\textsc{stirap}|=n\times (6 B_X)$, with $n=65$ an integer to maximize the coupling between states \ket{0} and \ket{2}.

To reduce the number of free parameters, we have taken  $\Om_{10}=\Om_{12}\equiv\tilde{\Om}$. 
The nonharmonic character of the rotational spectrum renders the detunings for transitions to and from \ket{A,J'=3} generally different from those to and from \ket{A,J'=1}, and the corresponding Rabi frequencies smaller. We model this, setting $\Om_{32}=\Om_{34}\equiv 0.7\tilde{\Om}$.
Similarly, we set the detunings $\Delta_1=\Delta_2 \equiv \Delta$.
As shown in Fig.~\ref{fig:evol5LS}, at the end of the pulse, the majority of population ($\approx92\%$) has been successfully transferred to state \ket{2}, while about 4\% has remained in \ket{1}, and barely another $4\%$ has leaked to \ket{4}.

\begin{table}[tb]
  \caption{Parameters used for the numerical evolution shown in Fig.~\ref{fig:evol5LS} of the five-level system described by Eq.~\eqref{eq:Ham5LS}, in units of $B_X$.}
  \label{tab:params5LS}
  \begin{tabular}{l|ccccccccc}
     Parameter &
     $\tilde\Om$ & $\Gamma_A$ & $\Delta$ & $t_0$ & $\tau$
        & $\om_1$ & $\om_2$ & $\phi_2-\phi_1$ & $\delta_\textsc{stirap} $ \\
     \hline 
     Value & $104$ & $10^{-4}$ & $\tilde\Om$ & 4 & 0.02
        & 3600 & 3598 & $1.34$~rad & $-0.016$ 
  \end{tabular}
\end{table}

In the case of narrowband lasers, the best transfer would occur at two-photon resonance, $\omega_1-\omega_2=E_2-E_0=6B_X$. 
The bandwidth of the ultrafast laser pulses certainly reduces the need to meet this condition. 
In fact, we have observed that tuning $\omega_1-\omega_2$ allows modification of the final population distribution between \ket{0} and \ket{2} without notably populating any undesired state. A similar tuning can be realized with the pulse phases, $\phi_{1,2}$, as well.

Translating the values given in Table~\ref{tab:params5LS} to the case of CaH (KRb), we find that pulses of duration $\tau=0.16~(18)$~ps, with a delay $\delta_\textsc{stirap}=-0.13~(-14)$~ps, focused to a beam waist of 100~\mumeter\ with an energy per pulse $E=76~(15)$~nJ, would do the job
, parameters which fall within experimental capabilities in current experiments~\cite{mizrahi2014,meijer2007}.

\section{Sensitivity to some experimental uncertainties}\label{sec:expt_uncertainties}

\subsection{Temperature and micromotion}
\label{ssec:microm}
The description of the system in terms of its normal modes relies on the assumption that ion and molecule displacements, $\delta x$, from their trap minima (defined taking into account the ion-dipole coupling, IDC) are small,
\begin{align}
  (\delta x)_{\mathrm{thermal}} \ll a_{\mathrm{HO}} \,,
\end{align}
with $a_{\mathrm{HO}}$ the corresponding harmonic oscillator length. Note that this requirement is less stringent than demanding that the motional state be the trap's ground state ($n_{\com,\str}=0$).
Now, as long as this harmonic approximation remains valid, the geometric character of the accumulated phase $\phi$ ensures that temperature should not
be a concern~\cite{leibfried03,garcia-ripoll03,garcia-ripoll05,murpetit2012}, as all thermal states will acquire the same (geometric) phase and provide the same signal.

Regarding micromotion, this will be a relevant source of uncertainty if the micromotion amplitude is of the order of, or larger than, the ion-dipole separation $z_0$, as this would mean that the ion would generally be misaligned with respect to the EDM axis, and a more elaborate study of the excitation modes would be necessary. Because of this, micromotion should be reduced to ensure that
\begin{align}
  (\delta x)_{\mathrm{micromotion}} \ll z_0 \,.
\end{align}
Experimentally, one can modify the ion-trap potentials to reduce micromotion amplitude or, more simply, keep $z_0$ larger than the expected micromotion amplitude; in both cases, micromotion effects can be neglected.


\subsection{Sensitivity to ion-dipole misalignment}\label{ssec:alignment}
With respect to the alignment of $\dipvec$ with the $z$ axis defined by the ion position, it can be seen~(cf.\ Fig.\ 2 in~\cite{murpetit2014apb})
that the effective potential where the ion is placed\textemdash formed by its own trapping fields plus the effect of IDC\textemdash is very smooth around the minimum. One needs a displacement along $x,y$ of magnitude $\gtrsim z_0/10$ to be sensitive to this source of error. When this
happens, the exact decoupling of the dynamics in the three directions $(x, y, z)$ will no longer hold. This can be seen as an effective coupling
between the \com\ and \str\ modes in $(x, y)$ and those along $z$. Such effects should become apparent only for times
$\gtrsim \om^{-1}\sqrt{z_0/(r_{\perp}\talpha)}$ which are $\gg \om^{-1}$
under the assumption above of small displacements, $r_{\perp}:=\sqrt{x^2+y^2} \ll z_0$.
Here, $\talpha=\hbar\alpha/(m\om^2 z_0^2)$ [cf.~\eqref{eq:rdip}] is a dimensionless parameter comparing the IDC energy with the trapping energy; typically, $\talpha\ll 1$ (cf.~\cite{murpetit2014apb}).


\bibliographystyle{apsrev4-1}
\bibliography{biblio-iondipol}

\end{document}